\documentclass[sigconf,screen]{acmart}

\AtBeginDocument{%
  \providecommand\BibTeX{{%
    \normalfont B\kern-0.5em{\scshape i\kern-0.25em b}\kern-0.8em\TeX}}}

\usepackage{booktabs}  
\usepackage{subcaption}
\usepackage[normalem]{ulem}
\usepackage{amsmath}
\usepackage{pifont}
\usepackage{xfrac}
\usepackage{algorithm}
\usepackage{algpseudocode}
\usepackage{url}
\usepackage{comment}
\usepackage{color}
\usepackage{fancyvrb}
\usepackage{balance}
\usepackage{listings}
\usepackage{multirow}
\usepackage{enumitem}
\usepackage{mathtools}
\usepackage{graphicx}
\usepackage{pbox}
\usepackage{makecell}
\usepackage{rotating}
\usepackage{adjustbox}
\usepackage{amsthm}
\usepackage{mathtools}
\usepackage{hyperref}

\usepackage[compatibility=false,font=small,labelfont=bf]{caption}

\newcommand{\cmark}{\ding{51}}

\newcommand{\jxperf}[0]{\mbox{\textsc{JXPerf}}}
\newcommand{\rwtrap}[0]{\mbox{\texttt{RW\_TRAP}}}
\newcommand{\wtrap}[0]{\mbox{\texttt{W\_TRAP}}}
\newcommand*\rot{\rotatebox{90}} 

\theoremstyle{}
\newtheorem{definition}{Definition}

\setcopyright{acmlicensed}

\begin{document}

\title{Pinpointing Performance Inefficiencies in Java}

\author{Pengfei Su}
\affiliation{
\institution{College of William \& Mary, USA}
}
\email{psu@email.wm.edu}

\author{Qingsen Wang}
\affiliation{
\institution{College of William \& Mary, USA}
}
\email{qwang06@email.wm.edu}

\author{Milind Chabbi}
\affiliation{
\institution{Scalable Machines Research, USA}
}
\email{milind@scalablemachines.org}

\author{Xu Liu}
\affiliation{
\institution{College of William \& Mary, USA}
}
\email{xl10@cs.wm.edu}

\begin{abstract}
Many performance inefficiencies such as inappropriate choice of algorithms or data structures, developers' inattention to performance, and missed compiler optimizations show up as \emph{wasteful} memory operations.
\emph{Wasteful} memory operations are those that produce/consume data to/from memory that may have been avoided.
We present, \jxperf{}, a lightweight performance analysis tool for pinpointing wasteful memory operations in Java programs.
Traditional byte-code instrumentation for such analysis (1) introduces prohibitive overheads and (2) misses inefficiencies in machine code generation.
\jxperf{} overcomes both of these problems.
\jxperf{} uses hardware performance monitoring units to sample memory locations accessed by a program and uses hardware debug registers to monitor subsequent accesses to the same memory.
The result is a lightweight measurement at machine-code level with attribution of inefficiencies to their provenance --- machine and source code within full calling contexts.
\jxperf{} introduces only 7\% runtime overhead and 7\% memory overhead making it useful in production.
Guided by \jxperf{}, we optimize several Java applications by improving code generation and choosing superior data structures and algorithms, which yield significant speedups.
 \end{abstract}

\copyrightyear{2019} 
\acmYear{2019} 
\acmConference[ESEC/FSE '19]{Proceedings of the 27th ACM Joint European Software Engineering Conference and Symposium on the Foundations of Software Engineering}{August 26--30, 2019}{Tallinn, Estonia}
\acmBooktitle{Proceedings of the 27th ACM Joint European Software Engineering Conference and Symposium on the Foundations of Software Engineering (ESEC/FSE '19), August 26--30, 2019, Tallinn, Estonia}
\acmPrice{15.00}
\acmDOI{10.1145/3338906.3338923}
\acmISBN{978-1-4503-5572-8/19/08}

\ccsdesc[500]{General and reference~Metrics}
\ccsdesc[500]{General and reference~Performance}
\ccsdesc[500]{Software and its engineering~Software maintenance tools}

\keywords{Java profiler, performance optimization, PMU, debug registers}

\maketitle

\section{Introduction}
\label{sec:introduction}

Managed languages, such as Java, have become increasingly popular in various domains, including web services, graphic interfaces, and mobile computing. 
Although managed languages significantly improve development velocity, they often suffer from worse performance compared with native languages. 
Being a step removed from the underlying hardware is one of the performance handicaps of programming in managed languages.
Despite their best efforts,  programmers, compilers, runtimes, and layers of libraries, can easily introduce various subtleties to find performance inefficiencies in managed program executions.
Such inefficiencies can easily go unnoticed (if not carefully and periodically monitored) or remain hard to diagnose (due to layers of abstraction and detachment from the underlying code generation, libraries, and runtimes).

Performance profiles abound in the Java world to aid developers to understand their program behavior.
Profiling for execution hotspots is the most popular one~\cite{perf,Levon:OProfile,jprofiler-WWW,yourkit-WWW,visualvm-WWW,oracle-studio-WWW}.
Hotspot analysis tools identify code regions that are frequently executed disregarding whether execution is efficient or inefficient (useful or wasteful) and hence significant burden is on the developer to make a judgement call on whether there is scope to optimize a hotspot.
Derived metrics such as  Cycles-Per-Instruction (CPI) or cache miss ratio offer slightly better intuition into hotspots but are still not a panacea.
Consider a loop repeatedly computing the exponential of the same number, which is obviously a wasteful work; the CPI metric simply acclaims such code with a low CPI value, which is considered a metric of goodness.

There is a need for tools that specifically pinpoint wasteful work and guide developers to focus on code regions where the optimizations are demanded.
Our observation, which is justified by myriad case studies in this paper, is that many inefficiencies show up as wasteful operations when inspected at the machine code level, and those which involve the memory subsystem are particularly egregious. 
Although this is not a new observation~\cite{Chabbi:2012:DTP:2259016.2259033,Wen:2017:REV:3037697.3037729,witch,Su:2019:RLS:3339505.3339628} in native languages, its application to Java code is new and the problem is particularly severe in managed languages. The following inefficiencies often show up as wasteful memory operations. 
\begin{description}[leftmargin=*]
\item[Algorithmic inefficiencies:] frequently performing a linear search shows up as frequently loading the same value from the same memory location. 
\item[Data structural inefficiencies:] using a dense array to store sparse data where the array is repeatedly reinitialized to store different data items shows up as frequent store-followed-by-store operations to the same memory location without an intervening load operation.
\item[\emph{Suboptimal code generations:}] missed inlining can show up as storing the same values to the same stack locations; missed scalar replacement shows up as loading the same value from the same, unmodified, memory location. 
\item[\emph{Developers' inattention to performance:}] recomputing the same method in successive loop iterations can show up as silent stores (consecutive writes of the same value to the same memory). For example, the Java implementation of NPB-3.0 benchmark IS~\cite{Bailey:1991:NPB:125826.125925} performs the expensive power method inside a loop and in each iteration, the power method pushes the same parameters on the same stack location. Interestingly, this inefficiency is absent in the C version of the code due to a careful implementation where the developer hoisted the power function out of the loop.
\end{description}

This list suffices to provide an intuition about the class of inefficiencies detectable by observing certain patterns of memory operations at runtime. 
Some recent Java profilers~\cite{Xu:2013:RTO:2509136.2509512,Nguyen:2013:CDC:2491411.2491416,Dhok:2016:DTG:2950290.2950360,toddler,ldoctor} identify inefficiencies of this form. 
However, these tools are based on exhaustive Java byte code instrumentation, which suffer from two drawbacks: (1) high (up to 200$\times$) runtime overhead, which prevents them from being used for production software; (2) missing insights into lower-level layers e.g., inefficiencies in machine code.

\begin{figure}
\begin{lstlisting}[firstnumber=142,language=java]
for (int bit = 0,dual = 1; bit < logn; bit++,dual *= 2) {
  ...
  for (int a = 1; a < dual; a++) {
    ...
    for (int b = 0; b < n; b += 2*dual) {
@\label{L3}@      int i = 2*b ;
@\label{L4}@      int j = 2*(b + dual);
      double z1_real = data[j];
      double z1_imag = data[j+1];
      double wd_real = w_real*z1_real - w_imag*z1_imag;
      double wd_imag = w_real*z1_imag + w_imag*z1_real;
@\label{L1}@@$\blacktriangleright$@     data[j] = data[i] - wd_real; 
      data[j+1] = data[i+1] - wd_imag;
@\label{L2}@@$\blacktriangleright$@     data[i] += wd_real; 
      data[i+1] += wd_imag;
    }}}
\end{lstlisting}
 \vspace{-0.3in}
\captionof{lstlisting}{Redundant memory loads in SPECjvm 2008 scimark.fft. {\tt data[i]} is loaded from memory twice in a single iteration whereas it is unmodified between these two loads.}
\vspace{-1em}
\label{lst:fft}
\end{figure}

\begin{figure}[t]
\begin{center}
\includegraphics[width=0.3\textwidth]{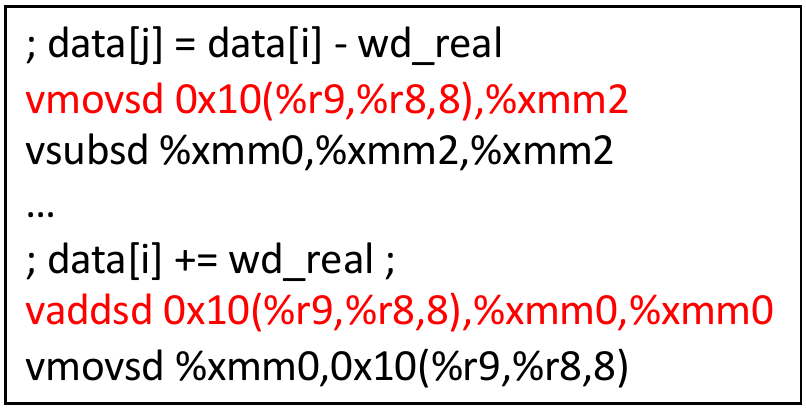}
\end{center}
\vspace{-0.15in}
\caption{The assembly code (at\&t style) of lines 153 and 155 in Listing~\ref{lst:fft}.} 
\vspace{-1em}
\label{fig:fft}
\end{figure}

\subsection{A Motivating Example}
\label{subsec:motivation}

\sloppy
Listing~\ref{lst:fft} shows a hot loop in a JIT-compiled (JITted) method (compiled with Oracle HotSpot JIT compiler) in SPECjvm2008 scimark.fft~\cite{SPEC:JVM2008}, a standard implementation of Fast Fourier Transforms.  
The JITted assembly code of the source code at lines \ref{L1} and \ref{L2} is shown in Figure~\ref{fig:fft}. 
Notice the two loads from the memory location \texttt{data[i]} (\texttt{0x10(\%r9,\%r8,8)}) --- once into  \texttt{xmm2} at line  \ref{L1} and then into  \texttt{xmm0} at line  \ref{L2}.
\texttt{data[i]} is unmodified between these two loads. 
Moreover, \texttt{i} and \texttt{j} differ by at least $2$ and never alias to the same memory location (see lines~\ref{L3} and~\ref{L4}).
Unfortunately, the code generation fails to exploit this aspect and trashes \texttt{xmm2} at line  \ref{L1}, which results in reloading \texttt{data[i]}  at line  \ref{L2}. 

With the knowledge of never-alias, we performed scalar replacement---placed \texttt{data[i]} in a temporary that eliminated the redundant load, which yielded a 1.13$\times$ speedup for the entire program.
Without access to the source code of the commercial Java runtime, we cannot definitively say whether the alias analysis missed the opportunity or the register allocator caused the suboptimal code generation, most likely the former.
However, it suffices to highlight the fact that observing the patterns of wasteful memory operations at the machine code level at runtime, divulges inefficiencies left out at various phases of transformation and allows us to peek into what ultimately executes.  
A more detailed analysis of this benchmark with the optimization guided by \jxperf{} follows in Section~\ref{subsec:fft}.

\subsection{Contribution Summary}

We propose \jxperf{} to complement existing Java profilers;  \jxperf{}  samples hardware performance counters and employs debug registers available in commodity CPUs to identify program inefficiencies that exhibit as wasteful memory operations at runtime. 
Two key differentiating aspects of \jxperf{} when compared to a large class of  hotspot profilers are its ability to (1) filter out and show code regions that are definitely involved in some kind of inefficiency at runtime (hotspot profilers cannot differentiate whether or not a code region is involved in any inefficiency), and (2) pinpoint the \emph{two parties} involved in wasteful work---e.g., the first instance of a memory access and a subsequent, unnecessary access of the same memory---which offer actionable insights (hotspot profilers are limited to showing only a single party).
Via \jxperf{}, we make the following contributions:

\begin{itemize}[leftmargin=*]
  \item  We show the design and implementation of a lightweight Java inefficiency profiler working on off-the-shelf Java virtual machine (JVM) with no byte code instrumentation to memory accesses.
  \item We demonstrate that \jxperf{} identifies inefficiency at machine code, which can be introduced by poor code generation, inappropriate data structures, or suboptimal algorithms.
  \item  We perform a thorough evaluation on \jxperf{} and show that \jxperf{}, with 7\% runtime overhead and 7\% memory overhead, is able to pinpoint inefficiencies in well-known Java benchmarks and real-world applications, which yield significant speedups after eliminating such inefficiencies.
\end{itemize}

\subsection{Paper Organization}
This paper is organized as follows. 
Section~\ref{sec:related} describes the related work and distinguishes \jxperf{}.
Section~\ref{sec:background} offers the background knowledge necessary to understand  \jxperf{}.
Section~\ref{sec:methodology} highlights the methodology we use to identify wasteful memory operations in Java programs.
Section~\ref{sec:design} depicts the design and implementation of \jxperf{}.
Section~\ref{sec:experiment} and~\ref{sec:use} evaluate \jxperf{} and show several case studies, respectively.
Section~\ref{sec:threats} discusses the threats to validity.
Section~\ref{sec:conclusion} presents our conclusions and future work.

\section{Related Work}
\label{sec:related}

There are a number of commercial and research Java profilers, most of which fall into the two categories: hotspot profilers and inefficiency profilers.
\paragraph{\textbf{\textit{Hotspot Profilers.}}}
Profilers such as Perf~\cite{perf}, Async-profiler~\cite{async-profiler-WWW}, Jprofiler~\cite{jprofiler-WWW}, YourKit~\cite{yourkit-WWW}, VisualVM~\cite{visualvm-WWW}, Oracle Developer Studio Performance Analyzer~\cite{oracle-studio-WWW}, and IBM Health Center~\cite{healthCenter-WWW} pinpoint hotspots in Java programs. 
Most hotspot profilers incur low overhead because they use interrupt-based sampling techniques supported by Performance Monitoring Units (PMUs) or OS timers. 
Hotspot profilers are able to identify code sections that account for a large number of CPU cycles, cache misses, branch mispredictions, heap memory usage, or floating point operations. 
While hotspot profilers are indispensable, they do not tell whether a resource is being used in a fruitful manner. For instance, they cannot report repeated memory stores of the identical value or result-equivalent computations, which squander both memory bandwidth and processor functional units.

\paragraph{\textbf{\textit{Inefficiency Profilers.}}}
Toddler~\cite{toddler} detects repeated memory loads in nested loops. LDoctor~\cite{ldoctor} combines static analysis and dynamic sampling techniques to reduce Toddler's overhead. However, LDoctor detects inefficiencies in only a small number of suspicious loops instead of in the whole program.
Glider~\cite{Dhok:2016:DTG:2950290.2950360} generates tests that expose redundant operations in Java collection traversals. 
MemoizeIt~\cite{DellaToffola:2015:PPY:2814270.2814290} detects redundant computations by identifying methods that repeatedly perform identical computations and output identical results. 
Xu et al. employ various static and dynamic analysis techniques (e.g., points-to analysis, dynamic slicing) to detect memory bloat by identifying useless data copying~\cite{Xu:2009:GFP:1542476.1542523}, inefficiently-used containers~\cite{Xu:2010:DIC:1806596.1806616}, low-utility data structures~\cite{Xu:2010:FLD:1806596.1806617}, reusable data structures~\cite{Xu:2013:RTO:2509136.2509512} and cacheable data structures~\cite{Nguyen:2013:CDC:2491411.2491416}. 

Unlike hotspot profilers, these tools can pinpoint redundant operations that lead to resource wastage. 
\jxperf{} is also an inefficiency profiler. 
Unlike prior works, which use exhaustive byte code instrumentation, \jxperf{} exploits features available in hardware (performance counters and debug registers) that eliminate instrumentation and dramatically reduces tool overheads. 
\jxperf{} detects multiple kinds of wasteful memory access patterns.
Furthermore, \jxperf{} can be easily extended with additional memory access patterns for detecting other kinds of runtime inefficiencies.
Section~\ref{sec:experiment} details the comparison between \jxperf{} and the profilers based on exhaustive byte code instrumentation.

Remix~\cite{remix}, similar to \jxperf{}, also utilized PMU; while \jxperf{} identifies intra-thread inefficiencies, such as redundant/useless operations,  Remix dentifies false sharing across threads.


\section{Background}
\label{sec:background}

\paragraph{\textbf{\textit{Hardware Performance Monitoring Units (PMU)}}}
Modern CPUs expose programmable registers that count various hardware events such as memory loads, stores, CPU cycles, and many others. 
The registers can be configured in sampling mode: when a threshold number of hardware events elapse, PMUs trigger an overflow interrupt. 
A profiler is able to capture the interrupt as a signal, known as a sample, and attribute the metrics collected along with the sample to the execution context. PMUs are per CPU core and virtualized by the operating system (OS) for each thread.

Intel offers Precise Event-Based Sampling (PEBS)~\cite{pmu-WWW} in SandyBridge and following generations. 
PEBS provides the effective address (EA) at the time of sample when the sample is for a  memory access instruction such as a load or a store. 
This facility is often referred to as address sampling, which is a critical building block of \jxperf{}. 
Also, PEBS can capture the precise instruction pointer (IP) for the instruction resulting in the counter overflow. 
AMD Instruction-Based Sampling (IBS)~\cite{AMDIBS:07} and PowerPC Marked Events (MRK)~\cite{Srinivas:2011:IBMJ-Power7} offer similar capabilities.

\paragraph{\textbf{\textit{Hardware Debug Registers.}}}
Hardware debug registers~\cite{Johnson:1982:RAS:800050.801837, McLear:1982:GCD:800050.801833} trap the CPU execution for debugging when the program counter (PC) reaches an address (breakpoint) or an instruction accesses a designated address (watchpoint). One can program debug registers to trap on various conditions: accessing addresses, accessing widths (1, 2, 4, 8 bytes), and accessing types (\wtrap{} and \rwtrap{}). The number of hardware debug registers is limited; modern x86 processors have only four debug registers.

\paragraph{\textbf{\textit{Linux perf\_event.}}}
Linux offers a standard interface to program PMUs and debug registers via the \texttt{perf\_event\_open} system call~\cite{perfevent} and the associated \texttt{ioctl} calls. 
A watchpoint exception is a synchronous CPU trap caused when an instruction accesses a monitored address,
while a PMU sample is a CPU interrupt caused when an event counter overflows. 
Both PMU samples and watchpoint exceptions are handled via Linux signals.
The user code can (1) mmap a circular buffer to which the kernel keeps appending the PMU data on each sample and (2) extract the signal context on each debug register trap.

\paragraph{\textbf{\textit{Java Virtual Machine Tool Interface (JVMTI)}}} 
\sloppy
JVMTI~\cite{jvmti-WWW} is a native programming interface of the JVM. A JVMTI client can develop a debugger/profiler (aka JVMTI agent) in any C/C++ based native language to inspect the state and control the execution of JVM-based programs. JVMTI provides a number of event callbacks to capture JVM initialization and death, thread creation and destruction, method loading and unloading, garbage collection start and end, to name a few. User-defined functions are registered in these callbacks and invoked when the associated events happen. In addition, JVMTI maintains a variety of information for queries, such as the map from the machine code of each JITted method to byte code and source code,  and the call path for any given point during the execution. JVMTI is available in off-the-shelf Oracle HotSpot JVM.

\section{Methodology}
\label{sec:methodology}

In the context of this paper, we define the following three kinds of wasteful memory accesses. 

\begin{definition}[Dead store]
$S_1$ and $S_2$ are two successive memory stores to location $M$ ($S_1$ occurs before $S_2$). $S_1$ is a dead store iff there are no intervening memory loads from $M$ between $S_1$ and $S_2$. In such a case, we call $\langle S_1, S_2\rangle$ a dead store pair.
\end{definition}

\begin{definition}[Slient store]
A memory store $S_2$, storing a value $V_2$ to location $M$, is a silent store iff the previous memory store $S_1$  performed on $M$ stored a value $V_1$, and $V_1 = V_2$. In such a case, we call $\langle S_1, S_2\rangle$ a silent store pair.
\end{definition}

\begin{definition}[Silent load]
A memory load $L_2$, loading a value $V_2$ from location $M$ is a silent load iff the previous memory load $L_1$  performed on $M$ loaded a value $V_1$, and $V_1 = V_2$. In such a case, we call $\langle L_1, L_2\rangle$ a silent load pair.
\end{definition}

Silent stores and silent loads are value-aware inefficiencies whereas dead stores are value-agnostic ones. 
We perform precise equality check on integer values, and approximate equality check on floating-point (FP) values within a user-specified threshold of difference (1\% by default), given the fact that all FP numbers under the IEEE 754 format are approximately represented in the machine. For memory operations involved in the inefficiencies, we typically use their calling contexts instead of their effective addresses to represent them, which can facilitate optimization efforts. 

\begin{figure}[t]
\begin{center}
\includegraphics[width=0.48\textwidth]{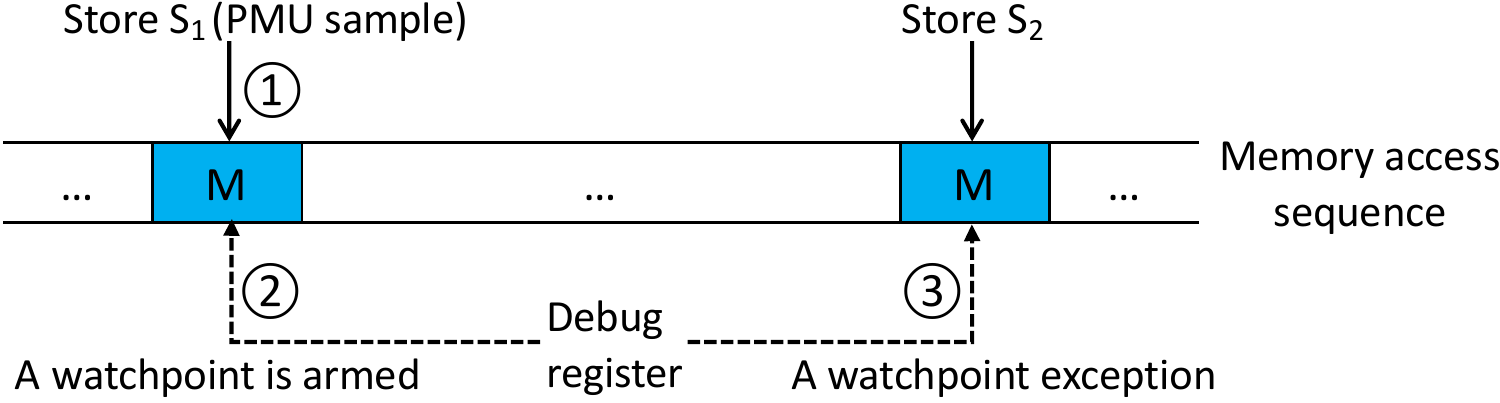}
\end{center}
\caption{\jxperf{}'s scheme for silent store detection. \textcircled{1} The PMU samples a memory store $S_1$ that touches location $M$. \textcircled{2} In the PMU sample handler, a debug register is armed to monitor subsequent access to $M$. \textcircled{3} The debug register traps on the next store $S_2$ to $M$. \textcircled{4} If $S_1$ and $S_2$ write the same values to $M$, \jxperf{} labels $S_2$ as a silent store and $\langle S_1, S_2\rangle$ as a silent store pair.}
\label{fig:scheme}
\vspace{-1em}
\end{figure}

Figure~\ref{fig:scheme} highlights the idea of \jxperf{} in detecting inefficiencies at runtime, exemplified with silent stores. 
PMU drives \jxperf{} by sampling precise memory stores.
For a sampled store operation, \jxperf{} records the effective address captured by the PMU, reads the value in this address,  and sets up a trap-on-store watchpoint on this address via the debug register. 
The subsequent store to the same effective address in the execution will trap. 
\jxperf{} captures the trap and checks the value at the effective address of the trap.
If the value remains unchanged between the two consecutive accesses, \jxperf{} reports a pair of silent stores. 
The watchpoint is disabled, and the execution continues as usual to detect more such instances.

\section{Design and Implementation}
\label{sec:design}

\begin{figure}[t]
\begin{center}
\includegraphics[width=0.33\textwidth]{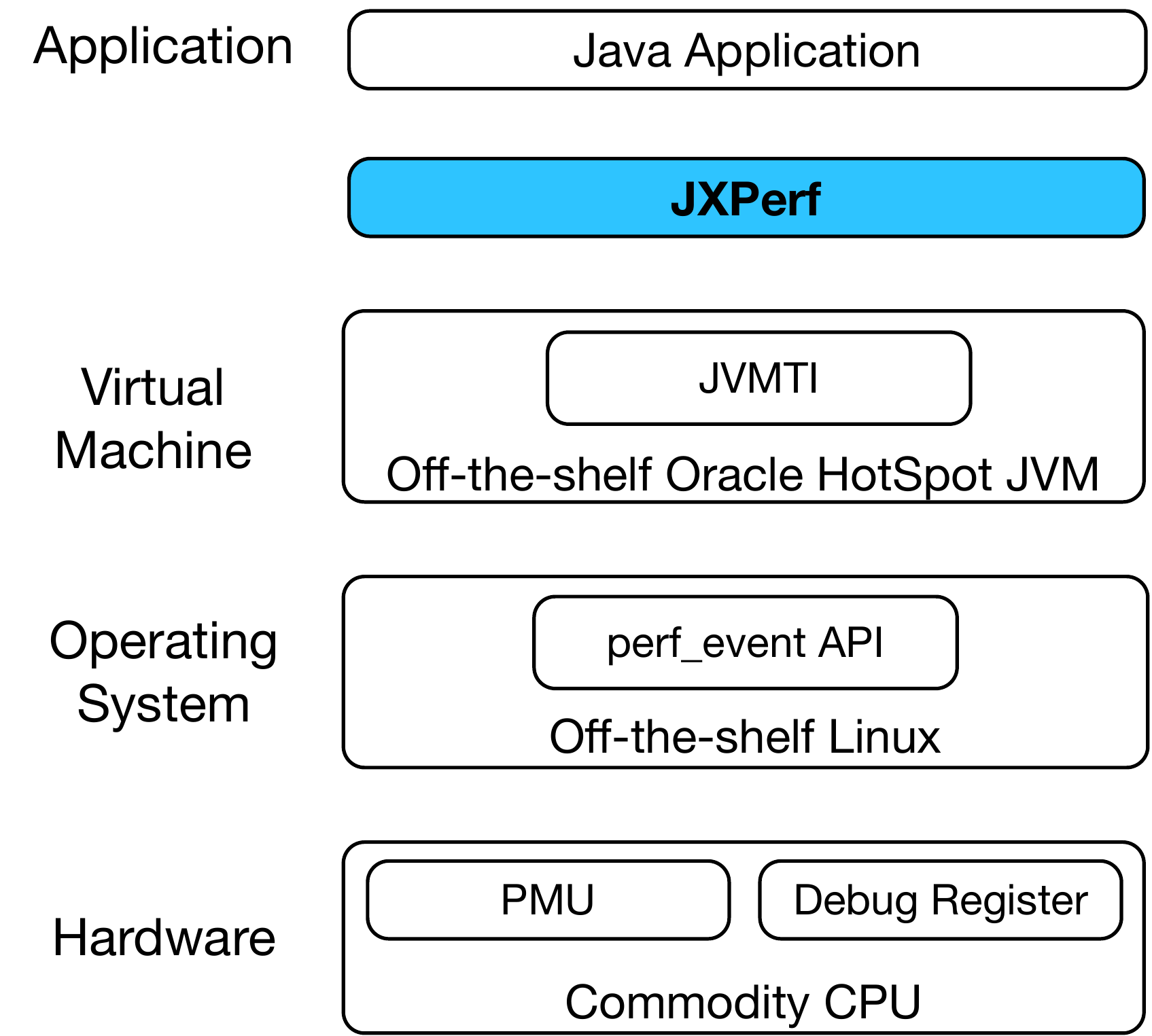}
\end{center}
\caption{Overview of \jxperf{} in the system stack.}
\label{fig:overview}
\vspace{-1em}
\end{figure}

Figure~\ref{fig:overview} overviews \jxperf{} in the entire system stack. 
\jxperf{} requires no modification to hardware (x86), OS (Linux), JVM (Oracle HotSpot), and monitored Java applications. In this section, we first describe the implementation details of \jxperf{} in identifying wasteful memory operations, then show how \jxperf{} addresses the challenges, and finally depict how \jxperf{} provides extra information to guide code optimization.  
\jxperf{} generates per-thread profiles and merges them to provide an aggregate view as the output.

\subsection{Lightweight Inefficiency Detection}
\label{sec:lightweight}


\paragraph{\textbf{\textit{Silent Store Detection.}}}
\begin{enumerate}
\item \jxperf{} subscribes to the precise PMU store event at the JVM initialization callbacks and sets up PMUs and debug registers for each thread via \texttt{perf\_event} API in the JVMTI thread creation callback. 
\item When a PMU counter overflows during the execution, it triggers an interrupt. 
 \jxperf{} captures the interrupt, constructs the calling context $C_1$ of the interrupt, and extracts the effective address $M$ and the value $V_1$ stored at $M$. 
\item  \jxperf{} sets a \wtrap{} (trap-on-store) watchpoint on $M$ and resumes the program execution. 
\item A subsequent store to $M$  triggers a trap.  \jxperf{} handles the trap signal, constructs the calling context $C_2$ of the trap, and inspects the value $V_2$ stored at $M$. 
\item  \jxperf{} compares $V_1$ and $V_2$.
If $V_1 = V_2$, a silent store is detected, and \jxperf{} labels the context pair $\langle C_1, C_2\rangle$ as an instance of silent store pair. 
\item \jxperf{}  disarms the debug register and resumes execution. 
\end{enumerate}

\paragraph{\textbf{\textit{Dead Store Detection.}}}
\jxperf{} subscribes to the precise PMU store event for dead store detection. When a PMU counter overflows, \jxperf{} constructs the calling context $C_1$ of the interrupt, extracts the effective address $M$,  sets a \rwtrap{} (load and store) watchpoint on $M$, and resumes program execution. 
When the subsequent access traps, \jxperf{} examines the access type (store or load).
If it is a store, \jxperf{} constructs the calling context $C_2$ of the trap and records the pair $\langle C_1, C_2\rangle$ as an instance of dead store pair. 
Otherwise, it is not a dead store.

\paragraph{\textbf{\textit{Silent Load Detection.}}}
The detection is similar to the silent store detection, except that \jxperf{} subscribes to the precise PMU load event and sets a \rwtrap{} watchpoint~\footnote{x86 debug registers do not offer trap-only-on-load facility.} to trap the subsequent access to the same memory address. 
If the watchpoint triggers on a load that reads the same value as the previous load from the same location, \jxperf{}  reports an instance of silent load pair.

The following metrics compute the fraction of wasteful memory operations in an execution:
\begin{eqnarray}
\scriptsize
\begin{aligned}
{\mathcal F}_{prog}^{DeadStore}=&{\sum_i\sum_j\text{Dead bytes stored in}\langle C_{i}, C_{j}\rangle \over \sum_i\sum_j\text{ Bytes stored in} \langle C_{i}, C_{j}\rangle} \\
{\mathcal F}_{prog}^{SilentStore}=&{\sum_i\sum_j\text{Silent bytes stored in}\langle C_{i}, C_{j}\rangle \over \sum_i\sum_j\text{ Bytes stored in} \langle C_{i}, C_{j}\rangle } \\
{\mathcal F}_{prog}^{SilentLoad}=&{\sum_i\sum_j\text{Silent bytes loaded from}\langle C_{i}, C_{j}\rangle \over \sum_i\sum_j\text{Bytes loaded from} \langle C_{i}, C_{j}\rangle}
\end{aligned}
\end{eqnarray}

Fraction of wasteful memory operations in a calling context pair is given as follows:
\begin{eqnarray}
\scriptsize
\begin{aligned}
{\mathcal F}_{\langle C_{watch}, C_{trap}\rangle}^{DeadStore}=&{\text{Dead bytes stored in}\langle C_{watch}, C_{trap}\rangle \over \sum_i\sum_j\text{Bytes stored in} \langle C_{i}, C_{j}\rangle}  \\
{\mathcal F}_{\langle C_{watch}, C_{trap}\rangle}^{SilentStore}=&{\text{Silent bytes stored in}\langle C_{watch}, C_{trap}\rangle \over \sum_i\sum_j\text{Bytes stored in} \langle C_{i}, C_{j}\rangle}  \\
{\mathcal F}_{\langle C_{watch}, C_{trap}\rangle}^{SilentLoad}=&{\text{Silent bytes loaded from}\langle C_{watch}, C_{trap}\rangle \over \sum_i\sum_j\text{Bytes loaded from} \langle C_{i}, C_{j}\rangle} 
\end{aligned}
\end{eqnarray}

\subsection{Limited Number of Debug Registers}
\label{sec:limited}

Hardware offers a small number of debug registers, which becomes a limitation if the PMU delivers a new sample before a previously set watchpoint traps.
To better understand the problem, consider the silent load example in Listing~\ref{lst:longDist}. 
Assume the loop indices \texttt{i} and \texttt{j}, and the scalars \texttt{sum1} and \texttt{sum2} are in registers.
Further assume the PMU is configured to deliver a sample every 1K memory loads and the number of debug register is only one. 
The first sample occurs in loop $i$ when accessing \texttt{array[1K]}, which results in setting a watchpoint to monitor the address of \texttt{array[1K]}. 
The second sample occurs when accessing \texttt{array[2K]}. 
Since the watchpoint armed at \texttt{array[1K]} is still active, we should either forgo monitoring it in favor of  \texttt{array[2K]} or ignore the new sample.
The former choice allows us to potentially detect a pair of silent loads separated by many intervening loads, and the latter choice allows us to detect a pair of silent loads separated by only a few intervening loads.
The option is not obvious without looking into the future.
A naive ``\emph{replace the oldest policy}'' is futile as it will not detect a single silent load in the above example.
Even a slightly smart \emph{exponential decay} strategy will not work because the survival probability of an old watchpoint will be minuscule over many samples.

\jxperf{} employs reservoir sampling~\cite{Vitter:1985:RSR:3147.3165,witch,rdx}, which uniformly chooses between old and new samples with no temporal bias. 
The first sampled address $M_1$, occupies the debug register with 1.0 probability. 
The second sampled address $M_2$, occupies the previously armed watchpoint with $\sfrac{1}{2}$ probability and retains $M_1$ with $\sfrac{1}{2}$ probability. 
The third sampled address $M_3$, either occupies the previously armed watchpoint with $\sfrac{1}{3}$ probability or retains it ($M_1$ or $M_2$) with $\sfrac{2}{3}$ probability. 
The $i^{th}$ sampled address $M_i$ since the last time a debug register was available, replaces the previously armed watchpoint with $\sfrac{1}{i}$ probability. 
The probability $P_k$ of monitoring any sampled address $M_k$, $1\le k \le i$, is the same ($\sfrac{1}{i}$), ensuring uniform sampling over time. 
When a watchpoint exception occurs, \jxperf{} disarms that watchpoint and resets its reservoir (replacement) probability to 1.0. 
Obviously, with this scheme \jxperf{} does not miss any sample if every watchpoint traps before being replaced. 

The scheme trivially extends to more number of debug registers, say $N \ge 1$.
\jxperf{} maintains an independent reservoir probability $P_\alpha$  for each debug register $\alpha$, ($1\le \alpha \le N$). 
On a PMU sample, if there is an available debug register, \jxperf{} arms it and decrements the reservoir probability of other already-armed debug registers; otherwise \jxperf{} visits each debug register $\alpha$ and attempts to replace it with the probability $P_\alpha$.
The process may succeed or fail in arming a debug register for a new sample, but it gives a new sample $N$ chances to remain in a system with $N$ watchpoints. Whether success or failure, $P_\alpha$ of each in-use debug register is updated after a sample. 
The order of visiting the debug registers is randomized for each sample to ensure fairness.
Notice that this scheme maintains only a count of previous samples (not an access log), which consumes $\mathcal{O}(1)$  memory.

\begin{figure}[t]
\begin{lstlisting}[firstnumber=1,language=java]
for (int i = 1; i <= 10K; i++) sum1 += array[i];
for (int j = 1; j <= 10K; j++) sum2 += array[j]; // silent loads
\end{lstlisting}
\vspace{-0.3in}
\captionof{lstlisting}{Long-distance silent loads. All four watchpoints are armed in the first four samples taken in loop \texttt{i} when the sampling period is 1K memory loads. Naively replacing the oldest watchpoint will not trigger a single watchpoint owing to many samples taken in loop \texttt{i} before reaching loop \texttt{j}. \jxperf{} employs the reservoir sampling to ensure each sample equal probability to survive.}
\label{lst:longDist}
\vspace{-1em}
\end{figure}

\subsection{Interference of the Garbage Collector}
Garbage collection (GC) can move live objects from one memory location to another memory location.
Without paying heed to GC events, \jxperf{} can introduce two kinds of errors: (1) it may erroneously attribute an instance of inefficiency (e.g., dead store) to a location that is in reality occupied by two different objects between two consecutive accesses by the same thread; (2) it may miss attributing an inefficiency metric to an object because it was moved from one memory location to another between two consecutive accesses by the same thread.

If \jxperf{} were able to query the garbage collector for moved objects or addresses, it could have avoided such errors, however, no such facility exists to the best of our knowledge in commercial JVMs.
\jxperf{}'s solution is to monitor accesses only between GC epochs. 
\jxperf{} captures the start and end points of GC by registering the JVMTI callbacks \texttt{GarbageCollectionStart} and \texttt{GarbageCollectionFinish} to demarcate epochs.
Watchpoints armed in an older epoch are not carried over to a new epoch: the first PMU sample or watchpoint trap that happens in a thread in a new epoch disarms all active watchpoints in that thread and begins afresh with a reservoir sampling probability of 1.0 for all debug registers for that thread.
Note that the GC thread is never monitored.
Typically, two consecutive accesses separated by a GC is infrequent; for example, the ratio of $\frac{\text{\# of GCs}}{\text{\# of PMU samples}}$ is 4.4e-5 in Dacapo-9.12-MR1-bach eclipse~\cite{Blackburn:2006:DBJ:1167473.1167488}.

\subsection {Attributing Measurement to Binary}
\label{sec:binary}
\jxperf{} uses Intel XED library~\cite{XED-WWW} for on-the-fly disassembly of JITed methods.
\jxperf{} retains the disassembly for post-mortem inspection if desired.
It also uses XED to determine whether a watchpoint trap was caused by a load or a store instruction.

A subtle implementation issue is in extracting the instruction that causes the watchpoint trap.
\jxperf{} uses the \texttt{perf\_event} API to register a \texttt{HW\_BREAKPOINT} perf event (watchpoint event) for a monitored address. 
Although the watchpoint causes a trap immediately after the instruction execution, the instruction pointer (IP) seen in the signal handler context (contextIP) is one ahead of the actual IP (trapIP) that triggers the trap.
In the x86 variable-length instruction set, it is nontrivial to derive the trapIP, even though it is just one instruction before the contextIP. 
The \texttt{HW\_BREAKPOINT} event in \texttt{perf\_event} is not a PMU event; hence, the Intel PEBS support, which otherwise provides the precise register state, is unavailable for a watchpoint. 
\jxperf{} disassembles every instruction from the method beginning till it reaches the IP just before the contextIP.
The expensive disassembly is amortized by caching results for subsequent traps that often happen at the same IP.
The caching is particularly important in methods with a large body; for example, when detecting silent loads in Dacapo-9.12-MR1-bach eclipse, without caching \jxperf{} introduces 4$\times$ runtime overhead.

\subsection{Attributing Measurement to Source Code}
\label{sec:cct}

\sloppy
Attributing runtime statistics to a flat profile (just an instruction) does not provide the full details needed for developer action.
For example, attributing inefficiencies to a common JDK method, e.g., \texttt{string.equals()}, offers little insight since \texttt{string.equals()} can be invoked from several places in a large code base; some invocations may not even be obvious to the user code. A detailed attribution demands associating profiles with the full calling context: \texttt{packageA.classB.methodC:line\#.->} \texttt{...->java.lang.String.equals():line\#.} Thus, \jxperf{} requires obtaining the calling context where a PMU sample occurs and the calling context where a watchpoint traps.

\paragraph{\textbf{\textit{Obtaining Calling Contexts without Safepoint Bias.}}}
Oracle JDK offers users two APIs to obtain calling contexts: officially documented \texttt{GetStackTrace()} and undocumented \texttt{AsyncGetCallTrace()}. Profilers that use \texttt{GetStackTrace()} suffer from the safepoint bias since JVM requires the program to reach a safepoint before collecting any calling context~\cite{Mytkowicz:2010:EAJ:1806596.1806618,Hofer:2014:FJP:2647508.2647509}. 
To avoid the bias, \jxperf{} employs \texttt{AsyncGetCallTrace()} to facilitate non-safepoint collection of calling contexts~\cite{AsyncGetCallTrace-WWW}. \texttt{AsyncGetCallTrace()} accepts \texttt{u\_context} obtained from the PMU interrupts or debug register traps as the input, and returns the method ID and byte code index (BCI) for each stack frame in the calling context. Method ID uniquely identifies distinct methods and distinct JITted instances of the same method (a single method may be JITted multiple times). With the method ID, \jxperf{} is able to obtain the associated class name and method name by querying JVM via JVMTI. To obtain the line number, \jxperf{} maintains a ``\texttt{BCI->line number}'' mapping table for each method instance by querying JVM via JVMTI API \texttt{GetLineNumberTable()}. As a result, for any given BCI, \jxperf{} returns its line number by looking up the mapping table.

\subsection{Post-mortem Analysis}
\label{sec:postmortem}
\jxperf{} produces per-thread profiles to minimize thread synchronization overhead during program execution.
We coalesce these per-thread profiles into a single profile in a post-mortem fashion.
The coalescing procedure follows the rule: two silent load (silent store or dead store) pairs from different threads are coalesced \emph{iff} they have the same loads (stores) in the same calling contexts. 
All metrics are also aggregated across threads. Typically, it takes less than one minute to merge all profiles according to our experiments.

\section{Evaluation}
\label{sec:experiment}

\sloppy
We evaluate \jxperf{} on an 18-core Intel Xeon E5-2699 v3 CPU of 2.30GHz frequency running Linux 4.8.0. The machine has 128GB main memory. \jxperf{} is built with Oracle JDK11 and complied with \texttt{gcc-5.4.1 -O3}. The Oracle HotSpot JVM is run in the server mode. 
\jxperf{} samples the PMU event \texttt{MEM\_UOPS\_RETIRED:ALL\_STORES} to detect dead stores and silent stores, and \texttt{MEM\_UOPS\_RETIRED:ALL\_LOADS} to detect silent loads. 

We evaluate \jxperf{} on three well-known benchmark suites---DaCapo 2006~\cite{Blackburn:2006:DBJ:1167473.1167488}, Dacapo-9.12-MR1-bach ~\cite{Blackburn:2006:DBJ:1167473.1167488} and ScalaBench~\cite{Sewe:2011:DCC:2048066.2048118}.
Additionally, we use two real-world performance bug datasets~\cite{toddler-bug-WWW, glider-WWW}.
All programs are built with Oracle JDK11 except DaCapo 2006 bloat, Dacapo-9.12-MR1-bach batik and eclipse, and ScalaBench actors with Oracle JDK8 due to the incompatibility. We apply the large input for DaCapo 2006, Dacapo-9.12-MR1-bach and ScalaBench, and the default inputs released with the remaining programs if not specified. Parallel programs, excluding threads used for the JIT compilation and GC, are run with four threads if users are allowed to specify the number of threads.

To deal with the impact of the non-deterministic execution (e.g., non-deterministic GC) of Java programs on experimental results, we refer to Georges et al.'s work~\cite{Georges:2007:SRJ:1297027.1297033} to use a confidence interval for the mean to report results. The confidence interval for the mean is computed based on the following formula where $n$ is the number of samples, $\overline{x}$ is the mean, $\sigma$ is the standard deviation and $z$ is a statistic determined by the confidence interval. In our experiments, we run each benchmark 30 times (i.e., $n=30$) and use a 95\% confidence interval (i.e., $z=1.96$).
\begin{eqnarray}
\small
\begin{aligned}
\overline{x} \pm z \times &{\sigma} \over {\sqrt{n}}
\end{aligned}
\end{eqnarray}

In the rest of this section, we first show the fraction of wasteful memory operations---dead stores, silent stores and silents loads on DaCapo 2006, Dacapo-9.12-MR1-bach, and ScalaBench benchmark suites at different sampling periods and different number of debug registers. We then evaluate the overhead of \jxperf{} on these benchmarks. We exclude three benchmarks---Dacapo-9.12-MR1-bach {\tt tradesoap}, {\tt tradebeans}, and {\tt tomcat}---from monitoring because of the huge variance in execution time of the native run ({\tt tradesoap} and {\tt tradebeans}) or runtime errors of the native run ({\tt tomcat}). Finally, we evaluate the effectiveness of \jxperf{} on the known performance bug datasets reported by existing tools.

\begin{figure}[t]
\begin{center}
\includegraphics[width=0.48\textwidth]{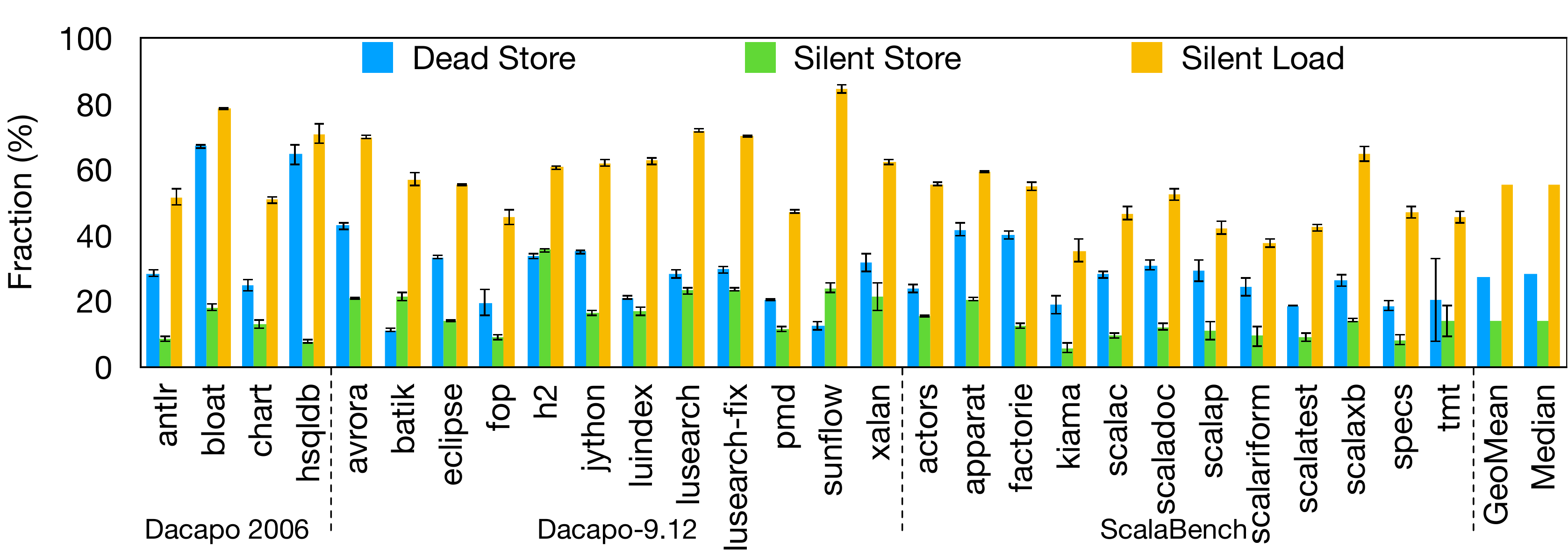}
\end{center}
\caption{Fraction of wasteful memory operations on DaCapo 2006, Dacapo-9.12-MR1-bach and ScalaBench benchmark suites at the sampling periods of 500K, 1M, 5M and 10M. The error bars are for different sampling periods.}
\label{fig:fraction-diff-rate}
\end{figure}

\begin{figure}[t]
\begin{center}
\includegraphics[width=0.48\textwidth]{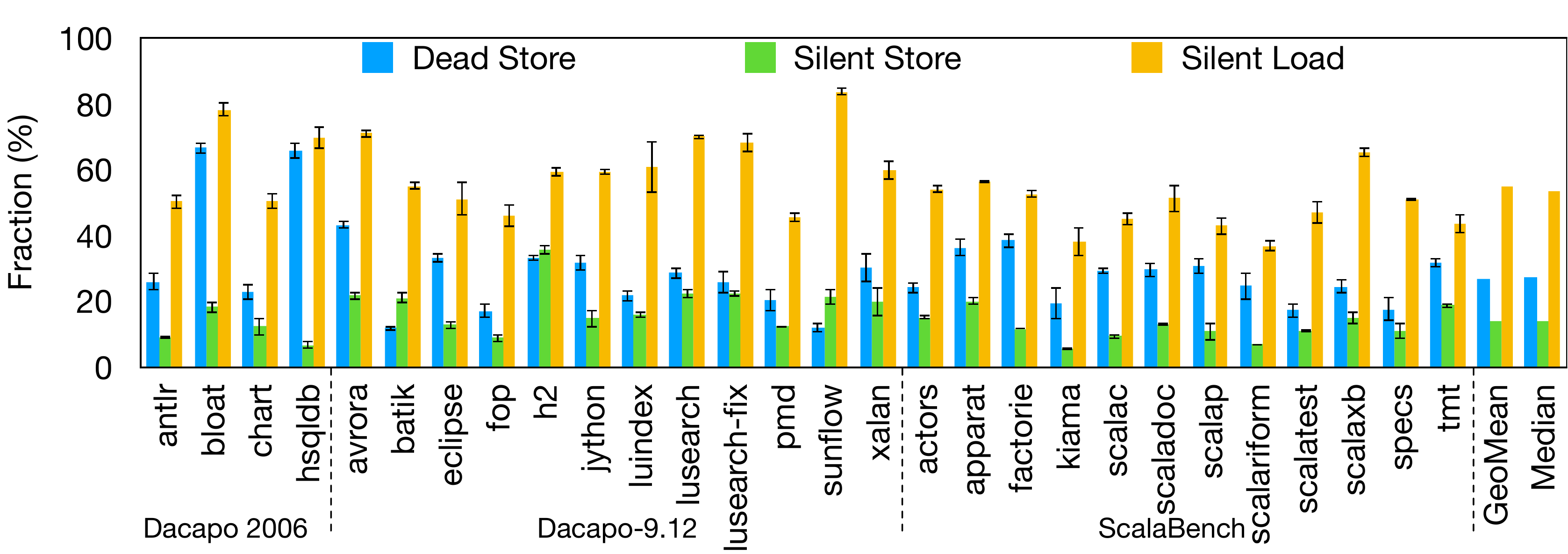}
\end{center}
\caption{Fraction of wasteful memory operations on DaCapo 2006, Dacapo-9.12-MR1-bach and ScalaBench benchmark suites by using different numbers of debug registers at the 5M sampling period. The error bars are for different number of debug registers.}
\label{fig:fraction-diff-db}
\end{figure}

\begin{table}[h]
\caption{Geometric mean and median of runtime slowdown ($\times$) and memory bloat ($\times$) of \jxperf{} at different sampling periods on DaCapo 2006, Dacapo-9.12-MR1-bach and ScalaBench benchmark suites (DS: dead store, SS: silent store, SL: silent load).}
\vspace{-.5em}
\begin{center}
\includegraphics[width=\linewidth]{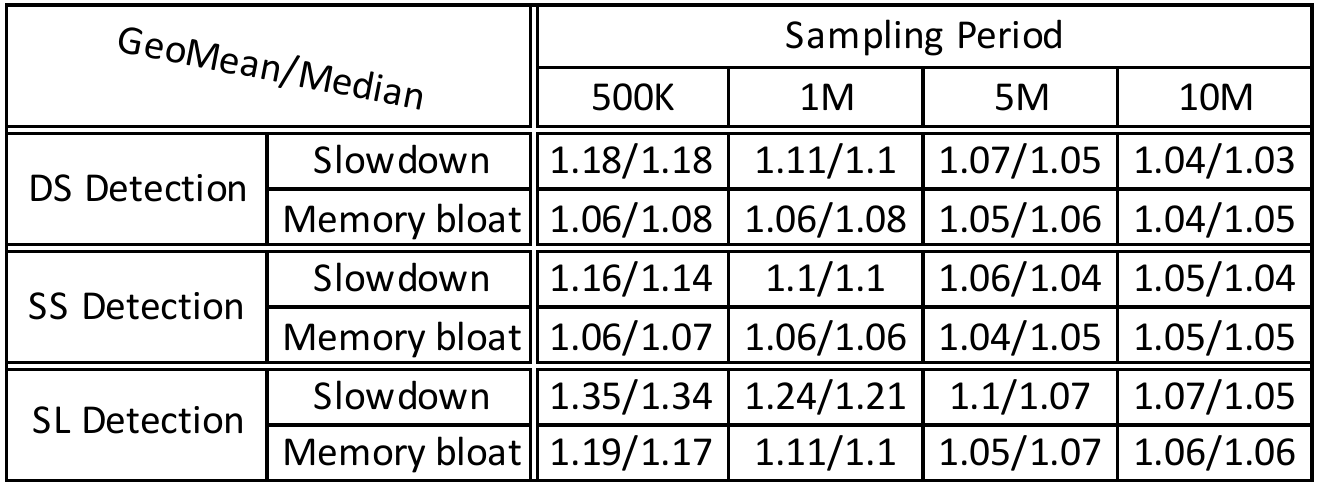}
\end{center}
\label{tab:overhead-all}
\end{table}

\begin{figure*}[t]
\begin{center}
\begin{subfigure}[b]{0.49\textwidth}
\includegraphics[width=\textwidth]{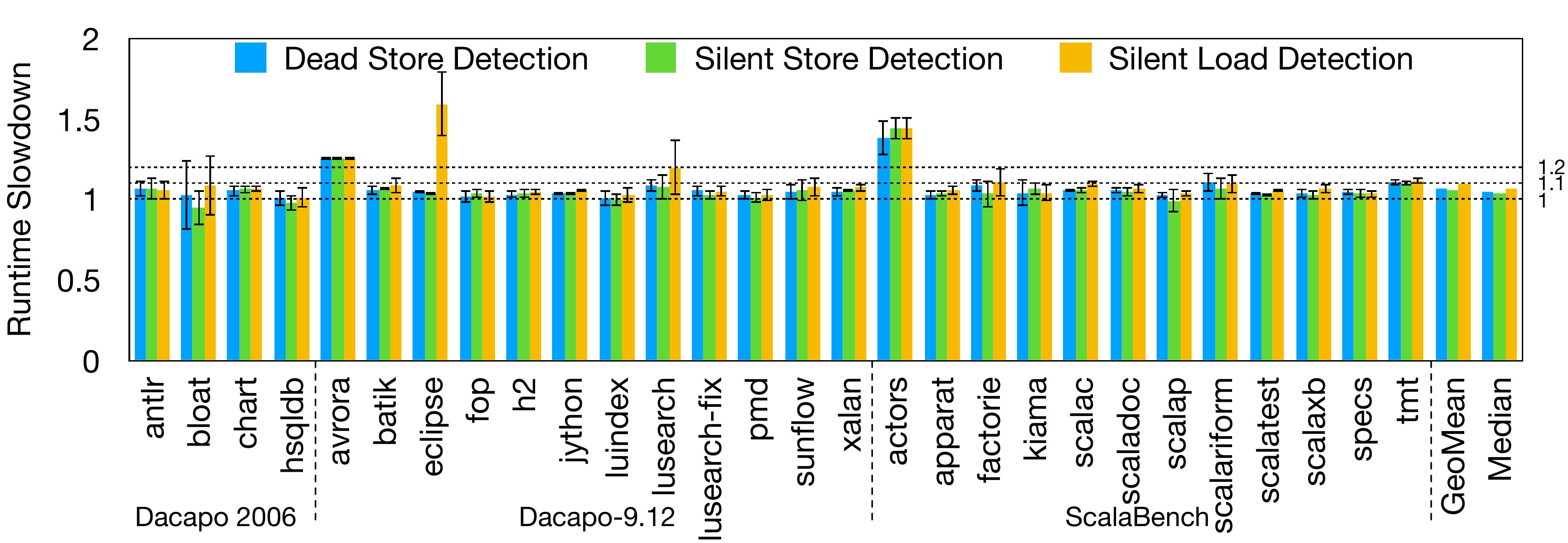}
\caption{Runtime slowdown.}
\label{fig:runtime-slowdown}
\end{subfigure}
~
\begin{subfigure}[b]{0.49\textwidth}
\includegraphics[width=\textwidth]{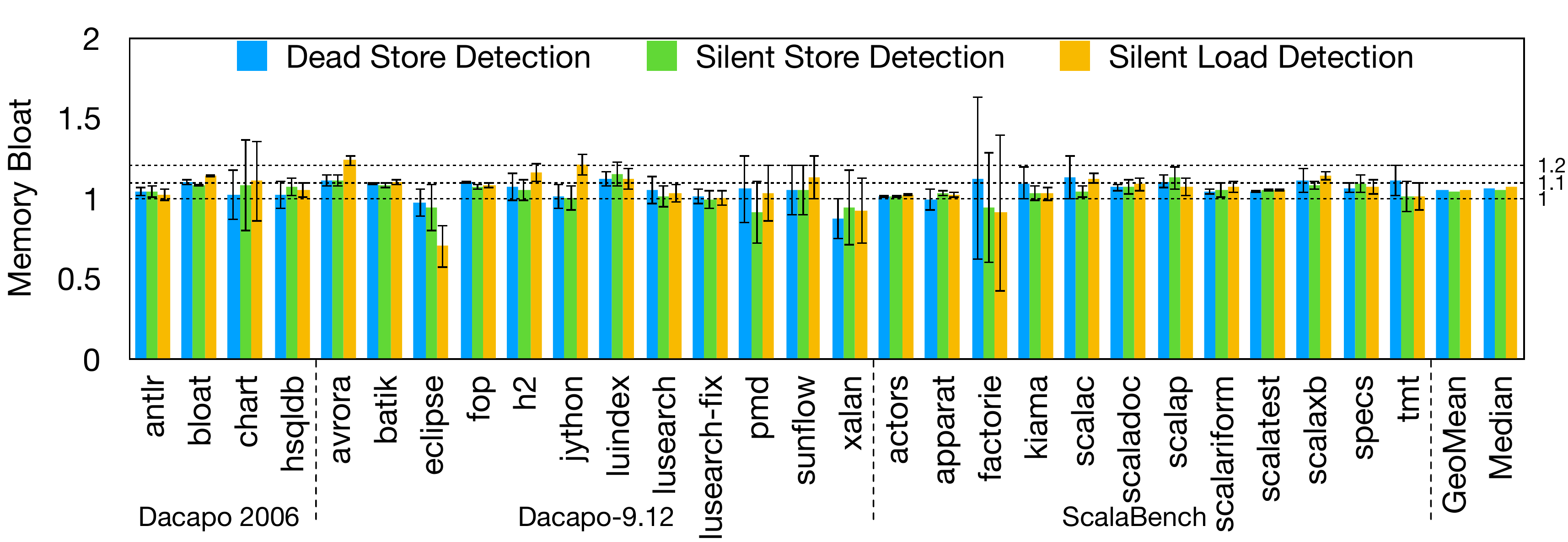}
\caption{Memory bloat.}
\label{fig:memory-bloat}
\end{subfigure}
\end{center}
\caption{Runtime slowdown ($\times$) and memory bloat ($\times$) of \jxperf{} at the 5M sampling period on DaCapo 2006, Dacapo-9.12-MR1-bach and ScalaBench benchmark suites.}
\vspace{-.8em}
\label{fig:overhead-5M}
\end{figure*}

\begin{table}[h]
\caption{Effectiveness of \jxperf{}. Toddler and Glider report 33 and 46 performance bugs from eight real-world applications, among which \jxperf{} succeeds in reproducing 31 and 44 bugs, respectively.}
\vspace{-.5em}
\begin{center}
\includegraphics[width=\linewidth]{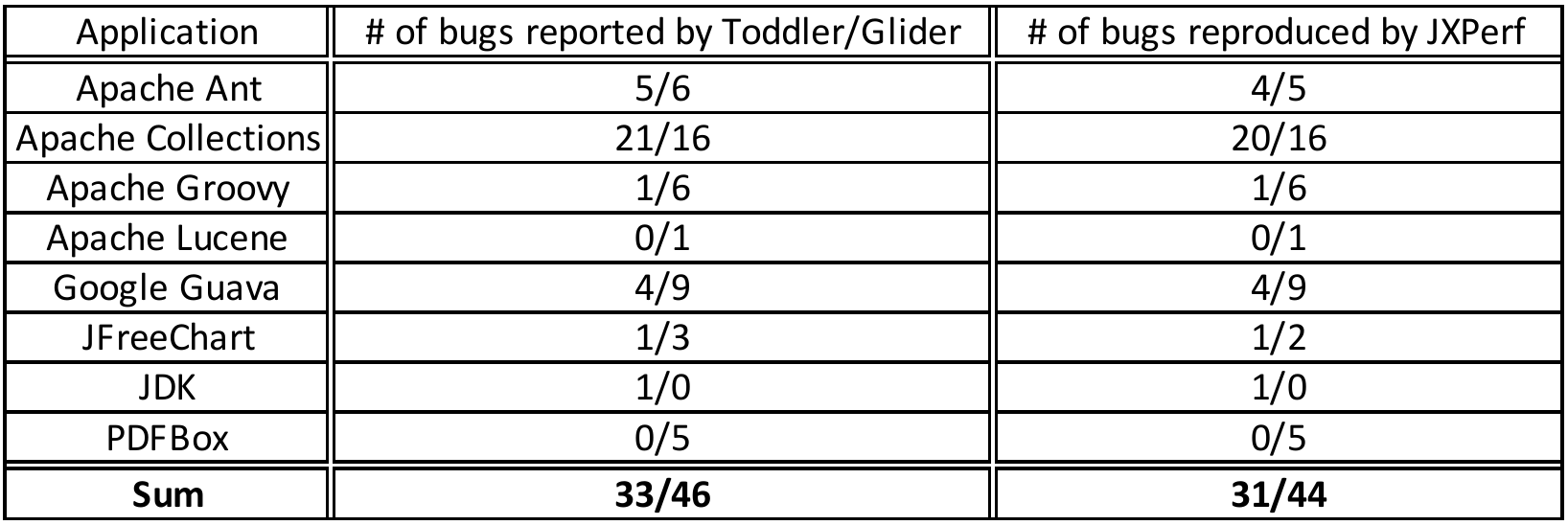}
\end{center}
\label{tab:bugs}
\vspace{-1em}
\end{table}

\paragraph{\textbf{\textit{Fraction of Wasteful Memory Operations.}}}
Figure~\ref{fig:fraction-diff-rate} shows the fraction of dead stores, silent stores, and silent loads on DaCapo 2006, Dacapo-9.12-MR1-bach, and ScalaBench benchmark suites at the sampling periods of 500K, 1M, 5M, and 10M. 
The following two takeaways are obvious: 
\begin{itemize}[leftmargin=*] 
\item The inefficiencies, such as dead stores, silent stores, and silent loads, pervasively exist in Java programs. 
\item The sampling period does not significantly impact the fraction of inefficiencies in Java programs. 
\end{itemize}

We further vary the number of debug registers from one to four to observe the variation in results at the same sampling period---5M, as shown in Figure~\ref{fig:fraction-diff-db}. We find the number of debug registers has minuscule impacts on the results except for a couple of short-running (e.g., $<$ 2s) benchmarks such as {\tt luindex} and {\tt kiama}, which validates the strength of the reservoir sampling.
We checked the top five inefficiency pairs and their percentage contributions and found negligible variance across different sampling periods and different number of debug registers.

\paragraph{\textbf{\textit{Overhead.}}}
Runtime slowdown (memory bloat) is measured as the ratio of the runtime (peak memory usage) of a benchmark with \jxperf{} enabled to the runtime (peak memory usage) of its native execution. Table~\ref{tab:overhead-all} shows the geometric mean and median of runtime slowdown and memory bloat at different sampling periods. When the sampling period increases (i.e., sampling rate decreases), the overhead drops as expected. We emperically find the 5M sampling period yields a good tradeoff between overhead and accuracy, which typically incurs 7\% runtime slowdown and 7\% memory bloat.

\sloppy
Figure~\ref{fig:overhead-5M} quantifies the overhead of \jxperf{} on each benchmark at the 5M sampling period. Silent load detection typically has a higher overhead than the other two because loads are more common than stores in a program execution. Moreover, \jxperf{} sets the \rwtrap{} (trap-only-on-load watchpoints are unavailable in x86 processors), which triggers an exception on both stores (ignored) and loads. From the program perspective, silent load detection for {\tt eclipse} incurs higher runtime overhead than others because it executes more load operations and has more methods of large size that require \jxperf{} to take more
efforts to correct the off-by-one error at each watchpoint trap. 
Furthermore, due to the non-deterministic behavior of GC, the peak memory usage for a couple of benchmarks with \jxperf{} enabled is less than the native run (e.g., {\tt eclipse}, {\tt xalan}) or varies significantly among different runs (e.g., {\tt factorie}).

\paragraph{\textbf{\textit{Effectiveness.}}}
We investigate the performance bugs reported by several state-of-the-art tools such as Toddler~\cite{toddler}, Clarity~\cite{Olivo:2015:SDA:2737924.2737966}, Glider~\cite{Dhok:2016:DTG:2950290.2950360}, and LDoctor~\cite{ldoctor}. Among them, the developers of Toddler and Glider share their bug datasets and test cases that expose the bugs online~\cite{toddler-bug-WWW, glider-WWW}. Therefore, we validate the effectiveness of \jxperf{} by checking whether the bugs reported by Toddler and Glider can also be identified by \jxperf{}. Toddler and Glider are both built atop Soot~\cite{soot-www} to identify a restricted class of performance issues: redundant operations involved in Java collection traversals, of which the symptom is silent loads. It is worth noting that the runtime overheads of Toddler and Glider are $\sim$16$\times$ and $\sim$150$\times$, respectively. 

Table~\ref{tab:bugs} shows the comparison results. 
Toddler reports 33 bugs (we exclude the bugs whose source codes or test cases are no longer available), 
among which \jxperf{} misses only two bugs: {\tt Apache Ant\#53637} and {\tt Apache Collections\#409}. 
Glider reports 46 bugs, among which \jxperf{} misses only two bugs: {\tt Apache Ant\#53637} and {\tt JFreeChart} (unknown bug ID).
Take Apache Collections\#588, one of the reported bugs, as an example to illustrate how \jxperf{} identifies it. Listing~\ref{lst:collection} shows the inefficient implementation of method \texttt{retainAll()} in Apache Collections\#588. \jxperf{} reports 49\% of silent loads are associated with method \texttt{contains()} at line  6 when the parameter \texttt{Collection} \texttt{coll} is of type list. For each element in \texttt{Iterator} \texttt{e}, \texttt{contains()} performs a linear search over \texttt{coll} to check whether \texttt{coll} contains this element. Consequently, elements in \texttt{coll} are repeatedly traversed whereas their values remain unchanged, which shows up as silent loads. 
Converting \texttt{coll} to a hash set is a superior choice of data structure that enables $\mathcal{O}(1)$ search algorithm and dramatically reduces both the number of loads and also the fraction of silent loads.

All the missed performance bugs fall into the same category: inefficiency observed in adjacent memory locations rather than the same memory location.
 We take {\tt Apache Ant\#53637} as an example to illustrate why \jxperf{} misses it. 
 The method ``\texttt{A.addAll(int index, Collection B)}'' in {\tt Ant\#53637} requires inserting elements of Collection {\tt A} one by one into the location ``index'' of Collection {\tt B}. 
 In each insertion, elements at and behind the location ``index'' of {\tt B} have to be shifted. 
 Consequently, elements in {\tt B} suffer from the repeated shifts. 
 The symptom of such inefficiency is that the same value is repeatedly loaded from adjacent memory locations.  
 \jxperf{} only identifies silent loads that repeatedly read the same value from the same memory location.
  \jxperf{} can be extended with a heuristic to record values at adjacent locations at the sample point and compare them in a watchpoint.
It is worth noting that inefficiencies identified by Toddler, Clarity, Glider, and  LDoctor are mostly related to load operations, whereas  \jxperf{} also identifies significant store-related inefficiencies.

\begin{figure}
\begin{lstlisting}[firstnumber=1,language=java]
public boolean retainAll(final Collection<?> coll) {
  if (coll != null) {
    boolean modified = false;
    final Iterator<E> e = iterator();
    while (e.hasNext()) {
@$\blacktriangleright$@    if (!coll.contains(e.next())) {
        e.remove();
        modified = true;
      }
    }
    return modified;
  } else return decorated().retainAll(null);
}
\end{lstlisting}
\vspace{-0.3in}
\captionof{lstlisting}{Inefficient implementation of method \texttt{retainAll()} in Apache Collections\#588. \jxperf{} reports that 49\% of silent loads are associated with method \texttt{contains()} at line  6 when the parameter \texttt{coll} is of type list.}
\vspace{-1em}
\label{lst:collection}
\end{figure}

\begin{table*}
\centering 
\begin{adjustbox}{width=0.9\textwidth}
\scriptsize
\begin{tabular}{|c|c||c|c|c||c|c|} 
\hline
\multicolumn{2}{|c||}{\multirow{2}{*}{Program}} & \multicolumn{3}{c||}  {Inefficiency} & \multicolumn{2}{c|} {Optimization} \\ \cline{3-7}
\multicolumn{2}{|c||}{} & Code & Type & Root cause & Patch & Speedup ($\times$) \\  \hline
\hline 
\multirow{4}{*}{\rot{\makecell{Macro \\benchmark}}} & \cmark SPECjvm2008 scimark.fft & FFT.java:loop(153-156) & SL& Poor binary code generation & Scalar replacement & 1.13$\pm$0.02 \\ \cline{2-7}
& \cmark NPB-3.0 IS & Random.java: randlc & SS & Redundant method invocations & Reusing the previous result& 1.89$\pm$0.04 \\ \cline{2-7}
& \cmark Grande-2.0 Euler & \makecell{Tunnel.java:calculateR \\Tunnel.java:calculateDamping} & DS &  Poor binary code generation & Scalar replacement & 1.1$\pm$0.02 \\ \hline
\multirow{3}{*}{\rot{\makecell{Real-world \\application}}} & \cmark SableCC-3.7 & \makecell{Grammar.java(15,16,64,65) \\LR0Collection.java(16,57,82,112) \\LR1Collection.java(16,17,27,28,33,34) \\LR0ItemSet.java(15,20,26) \\LR1ItemSet.java(15,20,26,124)} & SL & Poor data structure & Replacing TreeMap with LinkedHashMap & 3.08$\pm$0.32 \\ \cline{2-7}
& \cmark FindBugs-3.0.1 & Frame.java:copyFrom & DS & Inefficiently-used \texttt{ArrayList} & Improving \texttt{ArrayList} usage & 1.02$\pm$0.01 \\ \cline{2-7}
& \cmark Dacapo 2006 bloat & RegisterAllocator.java:loop(283) & DS & Useless value assignment in JDK & Removing the overpopulated containers & 1.35$\pm$0.05 \\ \cline{2-7}
& JFreeChart-1.0.19 & SegmentedTimeline.java:loop(1026) & SL & Poor linear search & Linear search with a break check & 1.64$\pm$0.04 \\ \hline
\multicolumn{4}{l}{{\vbox to 2ex{\vfil}}\scriptsize \cmark: newfound performance bugs via \jxperf{}.} \\
\multicolumn{4}{l}{{\vbox to 2ex{\vfil}}\scriptsize SS: silent store, DS: dead store, SL: silent load.} \\
\end{tabular}
\end{adjustbox}
\caption{Overview of performance improvement guided by \jxperf{}.}
\vspace{-1.5em} 	
\label{tab:case}
\end{table*}

\section{Case Studies}
\label{sec:use}

In addition to confirming the performance bugs reported by existing tools, we apply \jxperf{} on more benchmark suites---DaCapo 2006~\cite{Blackburn:2006:DBJ:1167473.1167488}, SPECjvm2008~\cite{SPEC:JVM2008}, NPB-3.0~\cite{Bailey:1991:NPB:125826.125925} and Grande-2.0~\cite{grande-WWW}, and real-world applications---SableCC-3.7~\cite{sablecc-WWW}, FindBugs-3.0.1~\cite{findbugs-WWW}, JFreeChart-1.0.19~\cite{jfreechart-WWW} to identify varieties of inefficiencies. 

Table~\ref{tab:case} summarizes the newly found performance bugs via \jxperf{} as well as previously found ones but with different insights provided by \jxperf{}.
All programs are built with Oracle JDK11 except Dacapo 2006 bloat and FindBugs-3.0.1, which are with Oracle JDK8.
We measure the performance of all programs in execution time except SPECjvm2008 scimark.fft, which is in throughput. 
We run each program 30 times and use a 95\% confidence interval for the mean speedup to report the performance improvement. 
In the rest of this section, we study each program shown in Table~\ref{tab:case}.

\subsection{SPECjvm2008 Scimark.fft:  Silent Loads}
\label{subsec:fft}

With the large input and four threads, \jxperf{} reports 33\% of memory loads are silent. The top two silent load pairs are attributed to lines 153 and 155, and lines 154 and 156 in Listing~\ref{lst:fft}, which account for 27\% of the total silent loads. They both suffer from the same performance issue: poor code generation detailed in Section~\ref{subsec:motivation}. We take lines 153 and 155 as an example to illustrate our optimization, of which the culprit calling contexts are shown in Figure~\ref{fig:fft-ctxt}.
We employ scalar replacement to eliminate such intra-iteration silent loads. In each iteration, we store the value of \texttt{data[i]} in a temporary before performing line 153, which enables \texttt{data[i]} to be loaded only once in each loop iteration. 
We also eliminate the silent loads between lines 154 and 156 using the same approach. They together eliminate 15\% of the total memory loads and yield a (1.13$\pm$0.02)$\times$ speedup for the entire program.

\begin{figure}[t]
\begin{minipage}{\linewidth}
\tiny
 \begin{Verbatim}[commandchars=\\\{\}]
--------------------------------------------------------------------------------
spec.harness.BenchmarkThread.run(BenchmarkThread.java:59)
 spec.harness.BenchmarkThread.executeIteration(BenchmarkThread.java:82)
  spec.harness.BenchmarkThread.runLoop(BenchmarkThread.java:170)
   spec.benchmarks.scimark.fft.Main.harnessMain(Main.java:36)
    spec.benchmarks.scimark.fft.Main.runBenchmark(Main.java:27)
     spec.benchmarks.scimark.fft.FFT.main(FFT.java:89)
      spec.benchmarks.scimark.fft.FFT.run(FFT.java:246)
       spec.benchmarks.scimark.fft.FFT.measureFFT(FFT.java:231)
        spec.benchmarks.scimark.fft.FFT.test(FFT.java:70)
         spec.benchmarks.scimark.fft.FFT.inverse(FFT.java:52)
          \textbf{vmovsd 0x10(%r9,%r8,8),%xmm2:...transform_internal(FFT.java:153)}
*********************************REDUNDANT WITH*********************************
spec.harness.BenchmarkThread.run(BenchmarkThread.java:59)
 spec.harness.BenchmarkThread.executeIteration(BenchmarkThread.java:82)
  spec.harness.BenchmarkThread.runLoop(BenchmarkThread.java:170)
   spec.benchmarks.scimark.fft.Main.harnessMain(Main.java:36)
    spec.benchmarks.scimark.fft.Main.runBenchmark(Main.java:27)
     spec.benchmarks.scimark.fft.FFT.main(FFT.java:89)
      spec.benchmarks.scimark.fft.FFT.run(FFT.java:246)
       spec.benchmarks.scimark.fft.FFT.measureFFT(FFT.java:231)
        spec.benchmarks.scimark.fft.FFT.test(FFT.java:70)
         spec.benchmarks.scimark.fft.FFT.inverse(FFT.java:52)
           \textbf{vaddsd 0x10(%r9,%r8,8),%xmm0,%xmm0:...transform_internal(FFT.java:155)}
--------------------------------------------------------------------------------
\end{Verbatim}
\end{minipage}
\vspace{-3ex}
\caption{A silent load pair with full calling contexts reported by \jxperf{} in SPECjvm2008 scimark.fft.}
\vspace{-1em}
\label{fig:fft-ctxt}
\end{figure}

\subsection{Grande-2.0 Euler: Dead Stores}
\label{subsec:euler}

Euler~\cite{grande-WWW} employs a structured mesh to solve the time-dependent Euler equations. \jxperf{} identifies 46\% stores are dead. One of the top dead store pairs is associated with the variable \texttt{temp2.a} at lines 5 and 12 in Listing~\ref{lst:euler}, which appears in a loop nest (not shown). 
By inspecting the JITted assembly code shown in Figure~\ref{fig:euler}, we find the value of \texttt{temp2.a} computed at line 5 is held in a register, which is reused at line 7. However, the memory store to \texttt{temp2.a} at line 5 is not eliminated. As a result, the memory store to \texttt{temp2.a} at line 12 overwrites the previous memory store to \texttt{temp2.a} at line 5.
Although CPUs buffer stores, workloads with many store operations, such as Euler~\cite{grande-WWW}, can cause CPU stalls due to store buffers filling up~\cite{6844459}.

To eliminate the dead stores, we use a temporary to replace \texttt{temp2.a} at lines 5, 7, and 12 to avoid using \texttt{temp2.a}. \jxperf{} also identifies other dead store pairs with the same issue and guides the same optimization. Our optimization eliminates 59\% of total memory stores and yields a (1.1$\pm$0.02)$\times$ speedup.
Our optimization is safe because \texttt{temp2} is a local object defined in the method {\tt calculateDamping} (line 2) to store the intermediate results; the object it refers to is never referenced by any other variable.

\begin{figure}
\begin{lstlisting}[firstnumber=1,language=java]
private void calculateDamping(double localpg[][], Statevector localug[][]) {
  Statevector temp2 = new Statevector();
  if (j > 1 && j < jmax-1) {
    temp = localug[i][j+2].svect(localug[i][j-1]);
@$\blacktriangleright$@   temp2.a = 3.0*(localug[i][j].a-localug[i][j+1].a);
    ...
    scrap4.a = tempdouble*(temp.a+temp2.a);
  }
  ...
  if (j > 1 && j < jmax-1) {
    temp = localug[i][j+1].svect(localug[i][j-2]);
@$\blacktriangleright$@   temp2.a = 3.0*(localug[i][j-1].a-localug[i][j].a);
    ...
  }
  ...
}
\end{lstlisting}
\vspace{-0.3in}
\captionof{lstlisting}{Dead stores in Grande-2.0 Euler. Successive memory stores to \texttt{temp2.a} without an intervening memory load.}
\label{lst:euler}
\end{figure}

\begin{figure}[t]
\begin{center}
\includegraphics[width=0.32\textwidth]{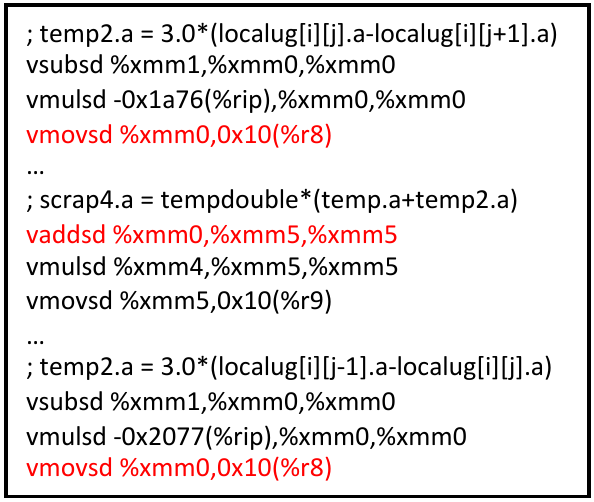}
\end{center}
\vspace{-0.15in}
\caption{The assembly code (at\&t style) of lines 5, 7 and 12 in Listing~\ref{lst:euler}.}
\label{fig:euler}
\end{figure}

\subsection{SableCC-3.7: Silent Loads}
\label{subsec:sable}

SableCC~\cite{sablecc-WWW} is a lexer and parser framework for compilers and interpreters. \jxperf{} profiles the latest stable version of SableCC by using the JDK7 grammar file as the input. \jxperf{} identifies that silent loads account for 94\% of the total memory loads and more than 80\% of silent loads are associated with method \texttt{put()} of the JDK \texttt{TreeMap} class. 
One of such top inefficiency pairs with calling contexts is shown in Figure~\ref{fig:sablecc}. 
The silent loads occur at line 568 in {\tt TreeMap.java}, whose source code is shown in Listing~\ref{lst:sablecc}. 
\texttt{TreeMap} is a Red-Black tree-based map where a \texttt{put} operation requires $\mathcal{O}(\log n)$ comparisons to insert an element. 
\texttt{put()} is frequently invoked to update the \texttt{TreeMap} during program execution.
Consequently, previously loaded elements in the \texttt{TreeMap} are often re-loaded to compare with new elements being inserted in different instances of \texttt{put()}, which shows up as silent loads.

By consulting the SableCC developers, we choose an alternative data structure.
We replace \texttt{TreeMap} with \texttt{LinkedHashMap} because (1) the linked list preserves ordering from one execution to another and (2) the hash table offers $\mathcal{O}(1)$ time complexity and obviously reduces the number of loads as well as the fraction of silent loads. 
We employ this transformation in five classes: \texttt{LR0ItemSet}, \texttt{LR1ItemSet}, \texttt{LR0Collection}, \texttt{LR1Collection}, and \texttt{Grammar}.
This optimization reduces the memory loads by 43\% and delivers a (3.08$\pm$0.32)$\times$ speedup to the entire program.

\begin{figure}[t]
\begin{minipage}{\linewidth}
\scriptsize
 \begin{Verbatim}[commandchars=\\\{\}]
-------------------------------------------------------------------
org.sablecc.sablecc.SableCC.main(SableCC.java:136)
  org.sablecc.sablecc.SableCC.processGrammar(SableCC.java:170)
             ...
    \textbf{mov 0x20(\%rbp),\%r10d: java.util.TreeMap.put(TreeMap.java:568)}
***************************REDUNDANT WITH***************************
org.sablecc.sablecc.SableCC.main(SableCC.java:136)
  org.sablecc.sablecc.SableCC.processGrammar(SableCC.java:170)
             ...
    \textbf{mov 0x20(\%rbp),\%r10d: java.util.TreeMap.put(TreeMap.java:568)}
--------------------------------------------------------------------
\end{Verbatim}
\end{minipage}
\vspace{-3ex}
\caption{A silent load pair reported by \jxperf{} in SableCC-3.7.}
\label{fig:sablecc}
\end{figure}

\begin{figure}
\begin{lstlisting}[firstnumber=561,language=java]
public V put(K key, V value) {
  Entry<K,V> t = root;
  ...
  do {
    parent = t;
    cmp = k.compareTo(t.key);
    if (cmp < 0)
@$\blacktriangleright$@     t = t.left;
    else if (cmp > 0)
      t = t.right;
    ...          
  } while (t != null);
  ...
}
\end{lstlisting}
\vspace{-0.3in}
\captionof{lstlisting}{Method put() of the JDK TreeMap class. A put operation requires $\mathcal{O}(\log n)$ comparisons to insert an element.}
\label{lst:sablecc}
\end{figure}

\subsection{NPB-3.0 IS: Silent Stores}
\label{subsec:is}

IS~\cite{Bailey:1991:NPB:125826.125925} sorts integers using the bucket sort. With the class B input and four threads, \jxperf{} pinpoints that 70\% of memory stores are silent, of which more than 50\% are associated with method \texttt{pow()} at lines 3-6 in Listing~\ref{lst:is}. We notice method \texttt{randlc()} is invoked in a hot loop (not shown) and the arguments passed to \texttt{pow()} are loop invariant. Across loop iterations, \texttt{pow()} pushes the same parameters on the same stack location, which shows up as silent stores.

To eliminate such redundant operations, we hoist the four calls to \texttt{pow()} outside of \texttt{randlc()} and memoize their return values in private class variables. 
\jxperf{} further identifies other code snippets having the same issue and guides the same optimization. 
These optimizations eliminate 96\% of memory stores and yield a (1.89$\pm$0.04)$\times$ speedup for the entire program.

\begin{figure}
\begin{lstlisting}[firstnumber=1,language=java]
public double randlc(double a) {
  double y[],r23,r46,t23,t46, ...;
@$\blacktriangleright$@ r23 = Math.pow(0.5,23); 
@$\blacktriangleright$@ r46 = Math.pow(r23, 2); 
@$\blacktriangleright$@ t23 = Math.pow(2.0,23);
@$\blacktriangleright$@ t46 = Math.pow(t23, 2);
  ...
}
\end{lstlisting}
\vspace{-0.3in}
\captionof{lstlisting}{Silent stores in NPB-3.0 IS. Method \texttt{pow()} repeatedly pushes the same parameters on the same stack across loop iterations.}
\label{lst:is}
\end{figure}

\subsection{Dacapo 2006 Bloat: Dead Stores}
\label{subsec:bloat}

Bloat~\cite{Blackburn:2006:DBJ:1167473.1167488} is a toolkit for analyzing and optimizing Java byte code. 
With the large input, \jxperf{} reports 78\% dead stores. 
More than 30\% of dead stores are attributed to the call site of method \texttt{addAll()} at lines 4 and 5 in Listing~\ref{lst:bloat}, where the program computes the union of \texttt{HashSet} \texttt{ig.succs(copy[0])} and \texttt{HashSet} \texttt{ig.succs(copy[1])}, and stores the result in \texttt{HashSet} ``\texttt{union}''. Guided by the culprit calling contexts, we notice the root cause of such dead stores is related to the field \texttt{current} of the JDK \texttt{HashMap} class, as shown in Listing~\ref{lst:jdk-hashmap}. Method \texttt{addAll()} invokes the method \texttt{nextNode()} of the \texttt{HashMap} class in a loop (not shown). In each iteration, the field \texttt{current} is overwritten with the newly inserted value, but never gets used during the execution, which shows up as dead stores.

With further code investigation, we find that \texttt{HashSet} ``\texttt{union}'' is created for only computing the size of the union of \texttt{ig.succs(copy[0])} and \texttt{ig.succs(copy[1])}, and elements in ``\texttt{union}'' are never used. Therefore, we can eliminate the dead stores by avoiding creating ``\texttt{union}''. We declare a counter variable to record the size of the union of \texttt{ig.succs(copy[0])} and \texttt{ig.succs(copy[1])}. The counter is initialized to the size of the larger one in \texttt{ig.succs(copy[0])} and \texttt{ig.succs(copy[1])}. Then we visit each element of the smaller one and check whether that element is already in the larger one. If not, the counter increments by 1. This optimization reduces 32\% of memory stores and yields a (1.35$\pm$0.05)$\times$ speedup for the entire program.

Yang et al.~\cite{Yang:2012:DAI:2338966.2336805} also identify the same optimization opportunity via the high-level container usage analysis, which is different from \jxperf{}'s binary-level inefficiency analysis.


\begin{figure}
\begin{lstlisting}[firstnumber=1,language=java]
union = new HashSet();
for (int i = 1; i < copies.size(); i++) {
  ...
@$\blacktriangleright$@ union.addAll(ig.succs(copy[0]));
@$\blacktriangleright$@ union.addAll(ig.succs(copy[1]));
  weight /= union.size();
  ...
}
\end{lstlisting}
\vspace{-0.3in}
\captionof{lstlisting}{Dead stores in Dacapo 2006 bloat. Useless value assignment in the JDK \texttt{HashMap} class leads to dead stores.}
\label{lst:bloat}
\end{figure}

\begin{figure}
\begin{lstlisting}[firstnumber=1,language=java]
final Node<K,V> nextNode() {
  Node<K,V>[] t;
  Node<K,V> e = next;
  ...
 @$\blacktriangleright$@if ((next=(current=e).next)==null&&(t=table)!=null) {
   do {} while (index<t.length&&(next=t[index++])==null);
  }
  return e;
}
\end{lstlisting}
\vspace{-0.3in}
\captionof{lstlisting}{Method {nextNode()} of the JDK \texttt{HashMap} class.}
\label{lst:jdk-hashmap}
\end{figure}

\subsection{FindBugs-3.0.1: Dead Stores}
\label{subsec:findbugs}

FindBugs~\cite{findbugs-WWW} is a static analysis tool for detecting security and performance bugs. We profile it using the JDK rt.jar as the input. \jxperf{} reports 47\% dead stores in this program. 
One of the top dead store pairs is attributed to the instance variable \texttt{ArrayList} \texttt{slotList} at lines 9 and 11 in Listing~\ref{lst:findbugs}. With an investigation into the implementation of the JDK \texttt{ArrayList} class, we find that the method {\tt clear()} assigns the \texttt{null} value to all elements in \texttt{slotList} and sets its size to zero instead of reclaiming the occupied space. When an element is inserted into \texttt{slotList} later by invoking the method \texttt{add()}, the \texttt{null} value at the given location of \texttt{slotList}, without any usage, is overwritten, which shows up as dead stores. 

We redesign the code to eliminate the dead stores, as shown in Listing~\ref{lst:findbugs-opt}. We first compare the size of ArrayList \texttt{slotList}, say $a$, with the size of \texttt{ArrayList} \texttt{other.slotList}, say $b$, to obtain the size of the smaller, say $min$. We then replace the first $min$ elements in \texttt{slotList} with the first $min$ elements in \texttt{other.slotList} (line 6) by invoking method \texttt{set()}. Finally, if $a$ > $min$, we invoke \texttt{clear()} to clear only the remaining elements in \texttt{slotList} (line 7); otherwise, we invoke \texttt{add()} to append the remaining elements in \texttt{other.slotList} to \texttt{slotList} (line 10). With this optimization, the total number of memory stores is reduced by 6\% and the entire program gains a (1.02$\pm$0.01)$\times$ speedup.

\begin{figure}
\begin{lstlisting}[firstnumber=1,language=java]
private final ArrayList<ValueType> slotList;
...
public void copyFrom(Frame<ValueType> other) {
  int size = slotList.size();
  if (size == other.slotList.size()) {
    for (int i = 0; i < size; i++)
      slotList.set(i, other.slotList.get(i));
  } else {
@$\blacktriangleright$@   slotList.clear();
    for (ValueType v : other.slotList)
@$\blacktriangleright$@     slotList.add(v);
  } 
  ...
}
\end{lstlisting}
\vspace{-0.3in}
\captionof{lstlisting}{Dead stores in FindBugs-3.0.1. Inefficiently-used \texttt{ArrayList} leads to dead stores.}
\label{lst:findbugs}
\end{figure}

\begin{figure}
\begin{lstlisting}[firstnumber=1,language=java]
public void copyFrom(Frame<ValueType> other) {
  int a = slotList.size();
  int b = other.slotList.size();
  int min = a > b ? b : a;
  for (int i = 0; i < min; i++) 
@$\blacktriangleright$@   slotList.set(i, other.slotList.get(i));
@$\blacktriangleright$@ if (a > min) slotList.subList(b,a).clear();
  else 
    for (int i = a; i < b; i++)
@$\blacktriangleright$@     slotList.add(other.slotList.get(i));
}
\end{lstlisting}
\vspace{-0.3in}
\captionof{lstlisting}{Optimizing the code in Listing~\ref{lst:findbugs} to eliminate dead stores.}
\label{lst:findbugs-opt}
\end{figure}

\subsection{JFreeChart-1.0.19: Silent Loads}
\label{subsec:jfreechart}

JFreeChart~\cite{jfreechart-WWW} is a chart library. \jxperf{} reports 90\% of memory loads are silent on profiling the built-in test case \texttt{SegmentedTimelineTest} and 30\% of silent loads are attributed to method \texttt{getExceptionSegmentCount()}, as shown in Listing~\ref{lst:jfreechart}. 
getExceptionSegmentCount() performs a linear search (line 7) over \texttt{ArrayList} \texttt{exceptionSegments} to count the number of segments that intersect a given segment [\texttt{fromMillisecond}, \texttt{toMillisecond}]. This linear search is called multiple times in a loop to become the performance bottleneck. The symptom of such inefficiency is silent loads, which is caused by the repeated loads of immutable \texttt{ArrayList} elements in different invocation instances of \texttt{getExceptionSegmentCount()}.

We notice that segments in \texttt{exceptionSegments} are stored in ascending order, that is, the end point of the segment \texttt{exceptionSegments.get(i)} $<$ the start point of the segment \texttt{exceptionSegments.get(j)} {\em iff} i $<$ j. Therefore, there is no need to traverse the remaining segments in \texttt{exceptionSegments} if the start point of the current segment is already greater than \texttt{toMillisecond}. 
With this optimization, we reduce the memory loads by 23\% and the entire program achieves a (1.64$\pm$0.04)$\times$ speedup. 

Nistor et al.~\cite{toddler} also identify the same performance issue with Toddler. However, their optimization~\cite{jfreechart-patch-WWW} guided by Toddler benefits the program only in two extreme situations: \texttt{toMillisecond} $<$ the start point of the first segment in \texttt{exceptionSegments} or \texttt{fromMillisecond} $>$ the end point of the last segment in \texttt{exceptionSegments}.
 
\begin{figure}
\begin{lstlisting}[firstnumber=1,language=java]
private List exceptionSegments = new ArrayList();
...
public long getExceptionSegmentCount(long fromMillisecond, long toMillisecond) {
  int n = 0;
  for (Iterator iter = this.exceptionSegments.iterator(); iter.hasNext();) {
    Segment segment = (Segment)iter.next();
@$\blacktriangleright$@   Segment intersection = segment.intersect(fromMillisecond, toMillisecond);
    if (intersection != null) {
      n += intersection.getSegmentCount();
    }}
  return (n);
}
\end{lstlisting}
 \vspace{-0.3in}
\captionof{lstlisting}{Silent loads in JFreeChart-1.0.19. Immutable \texttt{ArrayList} elements are repeatedly loaded from memory across invocation instances of method \texttt{getExceptionSegmentCount()}.}
\label{lst:jfreechart}
\end{figure}

\section{Threats to Validity}
\label{sec:threats}
\jxperf{} works on multi-threaded programs since PMUs and debug registers are virtualized by the OS for each program thread. \jxperf{} detects only intra-thread wasteful memory accesses and ignores inter-thread ones because a watchpoint traps \emph{iff} the same thread accesses the memory address the watchpoint is monitoring. 
Inter-thread access pattern detection is feasible but unwarranted for the class of inefficiencies pertinent to our current work.

Due to the intrinsic feature of sampling, \jxperf{} captures statistically significant memory accesses, but may miss some insignificant ones. 
Usually, optimizing such insignificant memory accesses yields trivial speedups.

Furthermore, not all reported wasteful memory operations need be eliminated. 
For example, compiler-savvy developers introduce a small number of silent loads or stores to make programs more regular for the purpose of vectorization, and security-savvy developers introduce a small number of dead stores to clear confidential data~\cite{yang:2017:ATC}. Developers' investigation or post-mortem analysis is necessary to make optimization choices. Only high-frequency inefficiencies are interesting; eliminating a long tail of insignificant inefficiencies that do not add up to a significant fraction is impractical and probably ineffective.

\section{Conclusions and Future Work}
\label{sec:conclusion}

Wasteful memory operations are often the symptoms of inappropriate data structure choice, suboptimal algorithm, and inefficient machine code generation. 
This paper presents \jxperf{}, a Java profiler that pinpoints performance inefficiencies arising from wasteful memory operations.
\jxperf{} samples CPU performance counters for addresses accessed by the program and uses hardware debug registers to monitor those addresses in an intelligent way. 
This hardware-assisted profiling avoids exhaustive byte code instrumentation and delivers a lightweight, effective tool, which does not compromise its ability to detect performance bugs. In fact, \jxperf{}'s ability to operate at the machine code level allows it to detect low-level code generation problems that are not apparent via byte code instrumentation.
\jxperf{} runs on off-the-shelf JVM, OS, and CPU, works on unmodified Java applications, and introduces only 7\% runtime overhead and 7\% memory overhead. 
Guided by \jxperf{}, we are able to optimize several benchmarks and real-world applications, yielding significant speedups.

As one of our future plans, we will extend \jxperf{} to other popular managed languages, such as Python and JavaScript, which recently employ JITters---PyPy~\cite{pypy-WWW} for Python and V8~\cite{v8-WWW} for JavaScript.

\section*{Acknowledgments}
We thank reviewers for their valuable comments. This work is supported by Google Faculty Research Award.

\bibliographystyle{ACM-Reference-Format}
\balance{}
\bibliography{references}

\end{document}